\documentclass{article}

\usepackage[english]{babel}

\usepackage[
  a4paper,
  top=2cm,
  bottom=2cm,
  left=3cm,
  right=3cm
]{geometry}

\usepackage{amsmath}
\usepackage{graphicx}
\usepackage[colorlinks=true, allcolors=blue]{hyperref}\usepackage{amsmath}
\usepackage{multirow}
\usepackage{graphicx}
\usepackage{threeparttable}
\usepackage{titlesec}
\usepackage{lmodern}
\usepackage{booktabs}
\usepackage{amsmath}
\usepackage{hyperref}
\usepackage{rotating}   % add this in your preamble if not already included
\usepackage{graphicx}
\usepackage{subcaption}   % <-- add this in the preamble
\usepackage[comma,authoryear]{natbib}
\usepackage{multibib}

\usepackage[bottom]{footmisc}
\title{\textbf{Tariffs and Labor Markets: The Employment Impact of the Recent Trade Conflict}\\
\large A Multiregional Input–Output Analysis}

\author{
\textbf{Christoph Ernst}\textsuperscript{1} \and
\textbf{Gabriel Michelena}\textsuperscript{2} \and
\textbf{Pablo Bertin}\textsuperscript{3}
}

\date{
\small
\textsuperscript{1}\,International Labour Organization (ILO), Geneva, Switzerland\\
\textsuperscript{2}\,Modelos Económicos de Simulación (MESi)-IIEP-UBA, Buenos Aires, Argentina\\
\textsuperscript{3}\,Modelos Económicos de Simulación (MESi)-IIEP-UBA, Buenos Aires, Argentina\\[0.8em]
 }

\begin{document}

\maketitle

\vspace{2.5cm}

\begin{abstract}

This paper assesses the global employment and trade effects of renewed tariff escalation following the reintroduction of the United States' ``America First'' strategy in 2025. Using a multiregional input–output (MRIO) framework integrated with a trade model, the analysis captures endogenous adjustments in both bilateral trade shares and final demand in response to changes in prices and competitiveness. 

\medskip

Three scenarios are simulated to reflect alternative configurations of trade policy: (i) existing tariffs without retaliation, (ii) updated tariffs including retaliatory measures, and (iii) a potential scenario characterized by de-escalation of the trade conflict. The results indicate that tariff increases generate widespread employment and export losses, with cumulative global job declines exceeding 23 million in the most adverse scenario. Informal and low-skilled workers bear the largest burden, accounting for more than 80 per cent of total employment losses, while high-income and upper middle-income countries experience significant contractions in export volumes. 

\medskip

The findings reveal that tariff shocks have regressive and asymmetric labour-market effects, amplifying existing inequalities across countries and worker groups. 

\end{abstract}

\vspace{1cm}


\vspace{2cm}

\small{Disclaimer: The views and opinions expressed in this article are those of the author and do not necessarily reflect the official policy or position of the institutions related to the authors.}

\newpage

\section{Introduction}

The international trade landscape experienced significant transformations between 2016 and 2018. The withdrawal of the United States from the Trans-Pacific Partnership (TPP), its exit from the Paris Agreement on climate change, and the termination of the Iran nuclear deal illustrate a reorientation of foreign policy. This shift was further evidenced by the suspension of negotiations on the Transatlantic Trade and Investment Partnership (TTIP) and the renegotiation of the North American Free Trade Agreement (NAFTA, later replaced by USMCA), both of which pointed to a more protectionist and less cooperative stance by the United States.
\medskip

The adoption of the ``America First'' strategy by Donald Trump in 2017 resulted in the institutionalisation of this unilateral approach to trade and the revision of key agreements. These changes significantly weakened the multilateral framework that had previously shaped international trade and marked a departure from the strong regionalist tendencies that characterised the 1980s and 1990s. The study explores the emergence of a new phase of trade tensions, characterised by an emphasis on tariffs and other trade barriers, particularly in the context of China. This development has the potential to reshape domestic and global production and trade structures \citep{boata_global_2024}. Despite the adoption of certain policies by the Biden administration, including the re-joining of the Paris Agreement and the refinancing of the World Health Organization (WHO), the emphasis of United States leadership on multilateral diplomacy could not be maintained beyond a single term of government. The change to a new administration led by Donald Trump, winner of the 2024 presidential election, marked a return to and deepening of the foreign policy adopted during the period 2017--2021.
\medskip

Considering forthcoming developments, recent declarations issued by the United States government indicate the beginning of a new phase marked by escalating trade tensions. This phenomenon can be interpreted as an indication that several countries are engaged in a contestation process regarding the prevailing global economic framework. This dynamic opens the door to a range of potential outcomes, from further fragmentation of global trade blocs to unexpected agreements among major players.
\medskip

The legal foundation for tariff increases is found in Section~232 of the Trade Expansion Act of 1962, which refers to national security, and Section~301 of the Trade Act of 1974. This legal basis was first invoked during the trade tensions that started in 2018 \citep{US_232_301}. These provisions have been invoked to justify the application of tariff measures against a broad set of trading partners, under arguments related to economic and national security and sovereignty. Consequently, such measures contribute to the deepening of geopolitical and economic fragmentation. However, as \cite{gopinath_changing_2025} observe, the levels of fragmentation today have not yet reached the characteristic of the Cold War period. However, stylized patterns of fragmentation in global trade and investment flows are becoming more prevalent.
\medskip

In this context, global and regional production chains are affected by changes in relative prices \citep{laborde2017}, which, in turn, influence the composition and demand for inputs from different trading partners, with direct repercussions on labor markets. Empirical evidence from the 2018 trade conflict reveals that tariff increases exert a detrimental effect on both output and employment, and that these effects are not homogeneous, as they depend on the sectors that benefit from protection and those adversely affected by the retaliatory measures of trading partners 
\citep{caliendo_trade_2021, fajgelbaum_economic_2022}. 
 
\medskip

Regarding this first round of trade tensions, \cite{laborde2017}, using a multiregional CGE model,  show that Mexico and China would suffer disproportionately, and that third countries can benefit as free riders. \cite{fajgelbaum_return_2020}, using a general equilibrium model with estimated trade elasticities, indicate that the 2018 tariffs led to full pass-through to U.S. import prices, large declines in bilateral trade, and significant welfare losses for U.S. consumers  (the aggregate real income loss was USD 7.2 billion, or 0.04 per cent of GDP).  \cite{fajgelbaum_economic_2022} provide a comprehensive synthesis of the effects of U.S.–China trade war, reviewing diverse empirical evidence confirming that the conflict reshaped global trade patterns, imposed substantial costs on U.S. consumers, and generated targeted retaliatory harm in politically sensitive U.S. regions. Finally, \cite{itakura2020} show that tariff escalation leads to large GDP losses of around 1.4 per cent in both China and the United States and broad sectoral contractions. A key innovation is the incorporation of agent-specific import demand to capture global value chains (GVCs), revealing that once GVC linkages are modelled explicitly, the negative effects become more widespread across countries and global GDP falls by roughly USD 450 billion. 
 \medskip

With respect to the trade tensions that began in 2025, there is a relative scarcity of research analysing their effects on labour markets. \citet{antoine_bouat_and_leysa_maty_sall_and_yu_zheng_towards_2025} employed a general equilibrium model to show that the loss of competitiveness experienced by firms has an adverse effect on employment. Furthermore, evidence indicates that in sectors more susceptible to trade dynamics, labour demand is adversely affected by rising import prices and the consequent decline in demand for these goods. \citet{cavalcanti_us-china_2025} examined the indirect effects of the trade conflict between China and United States on the Brazilian labour market, finding that China's retaliatory measures against the United States have a positive effect on formal employment in Brazil, while tariff increases on Chinese imports do not have a significant impact on the labour demand in Brazil.

\medskip

Based on the findings of preceding studies, the present document analyses and quantifies the effects of the trade tensions that have unfolded in 2025 on global employment. The paper makes a distinct contribution to the growing literature on trade policy shocks, global value chains, and labour-market adjustment by explicitly quantifying the employment effects of renewed tariff tensions through a price- and quantity-consistent multiregional input–output (MRIO) model.

\medskip

In this paper, the methodology of \citet{bardazzi_large-scale_2022} is adopted to carry out the simulations, integrating an MRIO accounting framework with simulated trade dynamics. Two features are of particular significance. First, trade shares are considered endogenous, while the estimation of bilateral import shares by sector is derived as a function of relative prices and final demand. In the event of fluctuations in prices or competitiveness, trade shares are reallocated, resulting in the redistribution of market shares among exporters. Second, the final demand is endogenous, with countries determining consumption and investment from income and price equations, thereby allowing nominal and real sides to interact. Higher prices can reduce real demand and shift its composition across origins.
\medskip

The results show that tariff shocks can generate asymmetric and non-zero-sum labour effects. In the event of multiple importers raising tariffs simultaneously, global demand experiences a contraction. At the same time, total employment levels typically decline, with the resulting job losses not offset by equivalent gains in other sectors. The tariff increases generate widespread employment and export losses, with cumulative global job declines exceeding 23 million in the most adverse scenario. Informal and low-skilled workers bear the largest burden, accounting for more than 80 per cent of total employment losses, while high-income and upper middle-income countries experience significant contractions in export volumes. 

\medskip

Regarding the contributions of existing literature, the present study is among the first to perform a systematic analysis of the 2025 phase of trade tensions on employment responses, employing an intersectoral and multilateral approach. The paper addresses this gap by examining the mechanism in which tariffs propagate through production chains and exert their influence on employment across multiple countries and industries. In addition, the contribution of this study lies in the integration of labour demand into the MRIO system. The employment model is formulated as a direct function of output, with sector-specific employment–output coefficients that respond to price and competitiveness shocks transmitted through the trade matrix. This approach captures the indirect employment effects along global value chains.

\medskip

The paper is structured as follows. Section~\ref{sec:chronology} outlines the chronology of recent trade tensions. Section~\ref{sec:empirical} describes the empirical strategy and data sources used in the simulations. Section~\ref{sec:scenarios} presents the policy scenarios, while Section~\ref{sec:results} discusses the main results, focusing on the effects of tariff escalation on exports and employment across countries, sectors, and labour groups. Finally, Section~\ref{sec:conclusion} concludes with key policy implications and suggestions for future research, emphasising the labour-market dimension of trade shocks and the need for policies that mitigate their regressive effects.

\clearpage

\section{The chronology of trade tensions}
\label{sec:chronology}

In the first week of February~2025, the United States announced new tariffs on Canada, Mexico, and China, marking the onset of an escalation in trade negotiations and tensions between the United States and several of its major partners. In a reciprocal action, the People's Republic of China imposed additional tariffs on the United States in the same week that it implemented its own tariff hikes. Subsequently, in the same month, the United States instigated an increase in tariffs on steel and aluminium by 25 per cent  for all trading partners.

\medskip

These announcements initiated a sequence of events characterized by further tariff hikes by the United States, rounds of negotiations, temporary suspensions of planned measures, and retaliatory actions by other countries. In response, the European Union, Japan, South Korea, Canada, and Mexico introduced countermeasures, raising tariffs on imports of United States origin. The most recent surge in tensions with China occurred in April, when tariffs exceeding 115 per cent  were imposed. On 2 April, the United States imposed new tariffs of 10 per cent  on all trading partners, in addition to further tariffs targeting countries with which it maintained substantial trade deficits. This action was referred as the ``Liberation Day.''

\medskip

As illustrated in Figure~1, the United States exhibits a merchandise trade deficit with all major regions of the world, including China. According to WITS/COMTRADE \footnote{\url{https://wits.worldbank.org/WITS/WITS/Restricted/Login.aspx},} it exceeded USD~1.6~trillion in 2023. The three regions showing the largest trade surpluses with the United States are East Asia and the Pacific, Europe and Central Asia, and China. These three regions account for more than 65 per cent  of the United States' global deficit. Consequently, these two regions and China have the highest \textit{ad valorem} tariffs, with Europe and Central Asia having the highest average tariffs at 21.8 per cent, while China and East Asia and the Pacific are below that with tariffs of 15.5 per cent  and 11.4 per cent, respectively. The countries of Africa and Latin America have the lowest tariffs, along with those that make up the rest of North America, due to the agreement signed with Canada and Mexico.

\begin{figure}[htbp]
\centering
\begin{subfigure}[t]{0.47\textwidth}
    \centering
    \includegraphics[width=\linewidth]{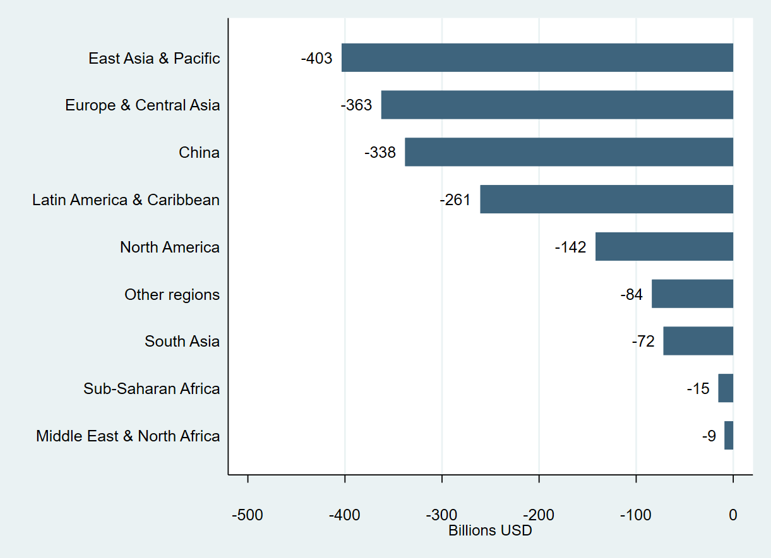}
    \caption{Trade balance (Billions USD) }
    \label{fig:scenario1}
\end{subfigure}
\hfill
\begin{subfigure}[t]{0.47\textwidth}
    \centering
    \includegraphics[width=\linewidth]{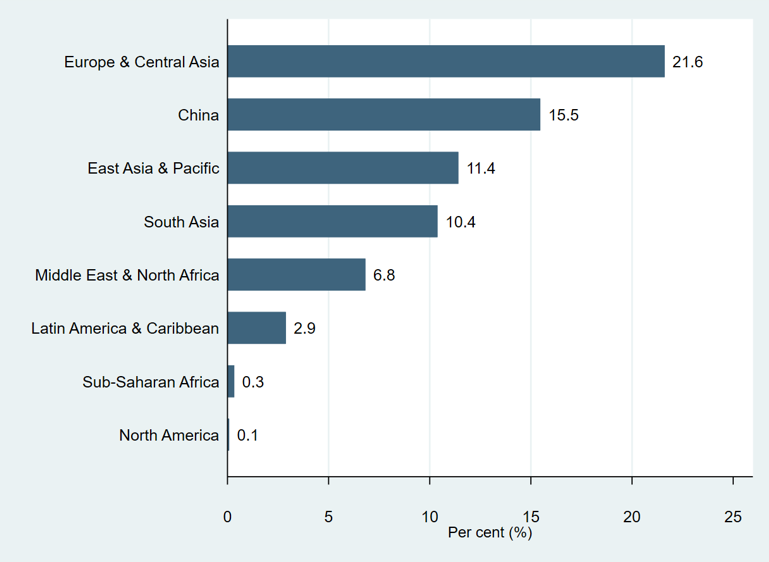}
    \caption{Effective Tariff applied (\%)}
    \label{fig:scenario2}
\end{subfigure}

\caption{Global employment and export impacts under alternative tariff escalation scenarios}
\label{fig:results}
\end{figure}

\medskip

\begin{table}[p]
\centering
\small
\setlength{\tabcolsep}{4pt} % reduce horizontal padding
\renewcommand{\arraystretch}{1.05} % tighter rows
\caption{Chronology of the main events in trade tensions from February to August 2025}
\label{tab:chronology}
\begin{threeparttable}
\begin{tabular}{p{1.6cm} p{1.1cm} p{5.2cm} p{5.2cm}}
\toprule
\textbf{Month} & \textbf{Week} & \textbf{United States} & \textbf{Retaliations / Agreements} \\
\midrule
\textbf{Feb.} & I & New tariffs on Canada, Mexico and China announced. Postpone Canada and Mexico tariff for 30 days. Tariffs on China take effect. & China applied new tariff to US. \\
 & II & 25\% tariffs on aluminium and steel imports implemented. & \\
\textbf{Mar.} & I & Delayed 25\% tariffs on Mexico and Canada take effect. Additional 10\% levy on Chinese imports announced. & Canada imposes 25\% retaliatory tariffs on USD 155 bn in imports, with USD 30 bn taking effect immediately. \\
 & II & Executive orders backtrack on 25\% tariffs on Mexico and Canada. Additional 25\% tariff on steel and aluminium imports from Canada. & EU and Canada retaliate: EU targets EUR 26 bn in United States goods from Apr 1; Canada imposes tariffs on CAD 30 bn (around USD 21 bn). \\
 & III & United States announces tariffs against countries that buy Venezuelan oil. & \\
 & IV & United States to impose 25\% tariff on foreign-made auto imports. & Japan, Korea, Europe and Mexico announce possible retaliation. \\
\textbf{Apr.} & I & Global tariffs of 10–50\%, incl. 10\% universal duty (Apr 5) and higher rates (Apr 9). & China announces 34\% tariff increase. \\
 & II & New 50\% tariff on Chinese goods (Apr 9); China’s rate raised to 145\%; 90-day pause for others at 10\%. & China raises retaliatory tariffs to 125\%, adding 50\% levy. \\
\textbf{May} & II & Tariffs on British cars and steel lowered; 10\% duties remain. China tariffs cut from 145\% to 30\% for 90 days. & China will cut tariffs from 125\% to 10\% and suspend non-tariff actions. \\
 & IV & 50\% tariff on EU goods delayed; talks extended to Jul 9; 50\% steel/aluminium tariff effective Jun 4. & \\
\textbf{Jun.} & I & New 50\% steel/aluminium tariffs effective; UK steel/aluminium remain at 25\%. & \\
\textbf{Jul.} & I & Trade deal with Vietnam: 20\% tariff on United States imports (was 46\%); 40\% tariff on goods transhipped via Vietnam. & \\
 & II & Deadline for “reciprocal” tariffs extended from Jul 9 to Aug 1. & \\
 & III & & Trade deal with Japan: 15\% tariffs on imports into US. \\
 & IV & US–EU tariff agreement: 15\% on most imports; India imports to face 25\% tariff (Aug 1). & EU removes tariffs on United States industrial products and gives preferential access to some agricultural goods. \\
\textbf{Aug.} & I & Tariffs on Indian imports increased over Russian oil purchases; 100\% tariff on non-US-made chips. & \\
 & II & Global tariffs take effect; 40\% tariff increase on Brazilian imports. & \\
 & III & Trump extends trade conflict truce with China for 90 days. & \\
\bottomrule
\end{tabular}
\begin{tablenotes}
\footnotesize
\item \textit{Source:} Authors’ elaboration based on USTR, \textit{Financial Times}, and ReedSmith.
\end{tablenotes}
\end{threeparttable}
\end{table}

\medskip

Beyond arguments related to bilateral trade imbalances and the illicit trafficking of fentanyl, other political and strategic considerations motivated the tariff reconfiguration. For instance, in the case of Brazil, the Trump administration imposed tariffs of 40 per cent  in August, ostensibly as a measure to incriminate former President Jair Bolsonaro, who was accused of orchestrating the dismissal of his successor, Luiz~Ignácio Lula da~Silva, when he assumed office. 

\medskip

Trade tensions with India also intensified, driven by India's purchases of Russian oil and gas, which represented more than 40 per cent  of Russia's exports to the rest of the world. At the same time, new tariff hikes were implemented in numerous sectors beyond steel and aluminium, frequently applied indiscriminately across trading partners. The pharmaceutical sector, along with crude and refined petroleum, semiconductors, storage battery components, copper, wood and aluminium inputs, and consumer goods such as smartphones, has been particularly impacted. The rationale behind this prioritisation is the reshoring of strategic manufacturing activities to specific regions, as well as the geopolitical, strategic, and economic significance of these areas.

\newpage

\section{Empirical Strategy}
\label{sec:empirical}

\subsection{Methodology}

To carry out the simulations, we extend the traditional MRIO framework \citep{miller_input-output_2009} to explicitly represent bilateral trade flows between countries or regions. Instead of aggregating international transactions into a single import vector for each destination, the model decomposes intermediate and final demand into flows by origin and destination. This structure allows the model to capture how each commodity produced in country~$o$ is used in country~$d$ across $n$ industries. Our model follows closely the Bilateral Trade Model of \citet{bardazzi_trade_2018,bardazzi_large-scale_2022}, although it shows some minor differences.

\medskip

For each destination region $d$, an inter-industry input–output matrix $Z_d$ describes the imported and domestic  intermediate goods used by each sector. The set of all $Z_d$ matrices forms the block-diagonal structure of the global input–output matrix $Z$, while international linkages are introduced through the bilateral supply matrices $SF_y$, which allocate, for each commodity $y$, the flows from all origin countries to all destinations. By stacking these $SF_y$ matrices, an allocation matrix $ALL$ is constructed, which defines the proportion of each product in the destination market that is supplied by each origin.

\begin{equation}
Z_d =
\begin{pmatrix}
z(d,1)_{1,1} & \cdots & z(d,1)_{1,n}\\
\vdots & \ddots & \vdots\\
z(d,n)_{n,1} & \cdots & z(d,n)_{n,n}
\end{pmatrix},
\qquad
SF_y =
\begin{pmatrix}
sf(y)_{1,1} & \cdots & sf(y)_{1,N}\\
\vdots & \ddots & \vdots\\
sf(y)_{N,1} & \cdots & sf(y)_{N,N}
\end{pmatrix}.
\label{eq:Z_SF}
\end{equation}

The MRIO system is represented by a global intermediate transactions matrix $Z$ of dimensions $Nn\times Nn$, where $N$ denotes the number of countries or regions and $n$ the number of industries in each. Each row and column of $Z$ therefore corresponds to a specific sector in a specific region. The overall matrix $Z$ takes a block-diagonal form:

\begin{equation}
Z =
\begin{pmatrix}
Z_1 & 0 & \cdots & 0\\
0 & Z_2 & \cdots & 0\\
\vdots & \vdots & \ddots & \vdots\\
0 & 0 & \cdots & Z_N
\end{pmatrix}.
\label{eq:Z_global}
\end{equation}

The vector of domestic final demand ($fd$) is constructed as the sum of total private consumption ($c$), government expenditure ($g$), and investment ($i$):

\begin{equation}
fd =
\begin{pmatrix}
fd_1\\ \vdots\\ fd_N
\end{pmatrix}
=
\begin{pmatrix}
c_1\\ \vdots\\ c_N
\end{pmatrix}
+
\begin{pmatrix}
g_1\\ \vdots\\ g_N
\end{pmatrix}
+
\begin{pmatrix}
i_1\\ \vdots\\ i_N
\end{pmatrix}.
\label{eq:fd}
\end{equation}

The export vector ($e$) consists of national export sub-vectors from each region:
\begin{equation}
e = 
\begin{pmatrix}
e_1\\ \vdots\\ e_N
\end{pmatrix}.
\label{eq:e}
\end{equation}
\medskip

Each block $sf(y)_{o,d}$ of the matrix ALL corresponds to a diagonal submatrix indicating a relocation of trade flows of commodity $y$ from origin $o$ to destination $d$. 

\begin{equation}
\resizebox{\textwidth}{!}{$
ALL=\left(\begin{matrix}\left.\ \begin{matrix}sf\left(1\right)_{1,1}&\cdots&0&\cdots&0\\\vdots&\ddots&\vdots&\ddots&\vdots\\0&\cdots&sf\left(y\right)_{1,1}&\cdots&0\\\vdots&\ddots&\vdots&\ddots&\vdots\\0&\cdots&0&\cdots&sf\left(n\right)_{1,1}\\\end{matrix}\right|\left.\ \begin{matrix}sf\left(1\right)_{1,2}&\cdots&0&\cdots&0\\\vdots&\ddots&\vdots&\ddots&\vdots\\0&\cdots&sf\left(y\right)_{1,2}&\cdots&0\\\vdots&\ddots&\vdots&\ddots&\vdots\\0&\cdots&0&\cdots&sf\left(n\right)_{1,2}\\\end{matrix}\right|\begin{matrix}\begin{matrix}\vdots\\\vdots\\\vdots\\\end{matrix}\\\vdots\\\vdots\\\end{matrix}\left|\begin{matrix}sf\left(1\right)_{1,N}&\cdots&0&\cdots&0\\\vdots&\ddots&\vdots&\ddots&\vdots\\0&\cdots&sf\left(y\right)_{1,N}&\cdots&0\\\vdots&\ddots&\vdots&\ddots&\vdots\\0&\cdots&0&\cdots&sf\left(n\right)_{1,N}\\\end{matrix}\right.\\\left.\ \begin{matrix}sf\left(1\right)_{2,1}&\cdots&0&\cdots&0\\\vdots&\ddots&\vdots&\ddots&\vdots\\0&\cdots&sf\left(y\right)_{2,1}&\cdots&0\\\vdots&\ddots&\vdots&\ddots&\vdots\\0&\cdots&0&\cdots&sf\left(n\right)_{2,1}\\\end{matrix}\right|\left.\ \begin{matrix}sf\left(1\right)_{2,2}&\cdots&0&\cdots&0\\\vdots&\ddots&\vdots&\ddots&\vdots\\0&\cdots&sf\left(y\right)_{2,2}&\cdots&0\\\vdots&\ddots&\vdots&\ddots&\vdots\\0&\cdots&0&\cdots&sf\left(n\right)_{2,2}\\\end{matrix}\right|\begin{matrix}\begin{matrix}\vdots\\\vdots\\\vdots\\\end{matrix}\\\vdots\\\vdots\\\end{matrix}\left|\begin{matrix}sf\left(1\right)_{2,N}&\cdots&0&\cdots&0\\\vdots&\ddots&\vdots&\ddots&\vdots\\0&\cdots&sf\left(y\right)_{2,N}&\cdots&0\\\vdots&\ddots&\vdots&\ddots&\vdots\\0&\cdots&0&\cdots&sf\left(n\right)_{2,N}\\\end{matrix}\right.\\\begin{matrix}\begin{matrix}\cdots&\cdots&\begin{matrix}\cdots&\cdots&\begin{matrix}\cdots&\cdots&\cdots\\\end{matrix}\\\end{matrix}\\\end{matrix}\\\left.\ \begin{matrix}sf\left(1\right)_{N,1}&\cdots&0&\cdots&0\\\vdots&\ddots&\vdots&\ddots&\vdots\\0&\cdots&sf\left(y\right)_{N,1}&\cdots&0\\\vdots&\ddots&\vdots&\ddots&\vdots\\0&\cdots&0&\cdots&sf\left(n\right)_{N,1}\\\end{matrix}\right|\left.\ \begin{matrix}sf\left(1\right)_{N,2}&\cdots&0&\cdots&0\\\vdots&\ddots&\vdots&\ddots&\vdots\\0&\cdots&sf\left(y\right)_{N,2}&\cdots&0\\\vdots&\ddots&\vdots&\ddots&\vdots\\0&\cdots&0&\cdots&sf\left(n\right)_{N,2}\\\end{matrix}\right|\begin{matrix}\begin{matrix}\vdots\\\vdots\\\vdots\\\end{matrix}\\\vdots\\\vdots\\\end{matrix}\left|\begin{matrix}sf\left(1\right)_{N,N}&\cdots&0&\cdots&0\\\vdots&\ddots&\vdots&\ddots&\vdots\\0&\cdots&sf\left(y\right)_{N,N}&\cdots&0\\\vdots&\ddots&\vdots&\ddots&\vdots\\0&\cdots&0&\cdots&sf\left(n\right)_{N,N}\\\end{matrix}\right.\\\end{matrix}\\\end{matrix}\right)
$}
\end{equation}

\medskip

Dividing each element of $ALL$ by its respective column total yields the matrix of allocation coefficients $T=[t_{o,d}]$, where each coefficient $t_{o,d}$ represents the share of supply of product $y$ in destination $d$ that originates from country $o$. By construction, $\sum_d t_{o,d}=1$ for all $d$, ensuring market balance.

\medskip

\begin{equation}
\resizebox{\textwidth}{!}{$
T=\left(\begin{matrix}\left.\ \begin{matrix}t\left(1\right)_{1,1}&\cdots&0&\cdots&0\\\vdots&\ddots&\vdots&\ddots&\vdots\\0&\cdots&t\left(y\right)_{1,1}&\cdots&0\\\vdots&\ddots&\vdots&\ddots&\vdots\\0&\cdots&0&\cdots&t\left(n\right)_{1,1}\\\end{matrix}\right|\left.\ \begin{matrix}t\left(1\right)_{1,2}&\cdots&0&\cdots&0\\\vdots&\ddots&\vdots&\ddots&\vdots\\0&\cdots&t\left(y\right)_{1,2}&\cdots&0\\\vdots&\ddots&\vdots&\ddots&\vdots\\0&\cdots&0&\cdots&t\left(n\right)_{1,2}\\\end{matrix}\right|\begin{matrix}\begin{matrix}\vdots\\\vdots\\\vdots\\\end{matrix}\\\vdots\\\vdots\\\end{matrix}\left|\begin{matrix}t\left(1\right)_{1,N}&\cdots&0&\cdots&0\\\vdots&\ddots&\vdots&\ddots&\vdots\\0&\cdots&t\left(y\right)_{1,N}&\cdots&0\\\vdots&\ddots&\vdots&\ddots&\vdots\\0&\cdots&0&\cdots&t\left(n\right)_{1,N}\\\end{matrix}\right.\\\left.\ \begin{matrix}t\left(1\right)_{2,1}&\cdots&0&\cdots&0\\\vdots&\ddots&\vdots&\ddots&\vdots\\0&\cdots&t\left(y\right)_{2,1}&\cdots&0\\\vdots&\ddots&\vdots&\ddots&\vdots\\0&\cdots&0&\cdots&t\left(n\right)_{2,1}\\\end{matrix}\right|\left.\ \begin{matrix}t\left(1\right)_{2,2}&\cdots&0&\cdots&0\\\vdots&\ddots&\vdots&\ddots&\vdots\\0&\cdots&t\left(y\right)_{2,2}&\cdots&0\\\vdots&\ddots&\vdots&\ddots&\vdots\\0&\cdots&0&\cdots&t\left(n\right)_{2,2}\\\end{matrix}\right|\begin{matrix}\begin{matrix}\vdots\\\vdots\\\vdots\\\end{matrix}\\\vdots\\\vdots\\\end{matrix}\left|\begin{matrix}t\left(1\right)_{2,N}&\cdots&0&\cdots&0\\\vdots&\ddots&\vdots&\ddots&\vdots\\0&\cdots&t\left(y\right)_{2,N}&\cdots&0\\\vdots&\ddots&\vdots&\ddots&\vdots\\0&\cdots&0&\cdots&t\left(n\right)_{2,N}\\\end{matrix}\right.\\\begin{matrix}\begin{matrix}\cdots&\cdots&\begin{matrix}\cdots&\cdots&\begin{matrix}\cdots&\cdots&\cdots\\\end{matrix}\\\end{matrix}\\\end{matrix}\\\left.\ \begin{matrix}t\left(1\right)_{N,1}&\cdots&0&\cdots&0\\\vdots&\ddots&\vdots&\ddots&\vdots\\0&\cdots&t\left(y\right)_{N,1}&\cdots&0\\\vdots&\ddots&\vdots&\ddots&\vdots\\0&\cdots&0&\cdots&t\left(n\right)_{N,1}\\\end{matrix}\right|\left.\ \begin{matrix}t\left(1\right)_{N,2}&\cdots&0&\cdots&0\\\vdots&\ddots&\vdots&\ddots&\vdots\\0&\cdots&t\left(y\right)_{N,2}&\cdots&0\\\vdots&\ddots&\vdots&\ddots&\vdots\\0&\cdots&0&\cdots&t\left(n\right)_{N,2}\\\end{matrix}\right|\begin{matrix}\begin{matrix}\vdots\\\vdots\\\vdots\\\end{matrix}\\\vdots\\\vdots\\\end{matrix}\left|\begin{matrix}t\left(1\right)_{N,N}&\cdots&0&\cdots&0\\\vdots&\ddots&\vdots&\ddots&\vdots\\0&\cdots&t\left(y\right)_{N,N}&\cdots&0\\\vdots&\ddots&\vdots&\ddots&\vdots\\0&\cdots&0&\cdots&t\left(n\right)_{N,N}\\\end{matrix}\right.\\\end{matrix}\\\end{matrix}\right)
$}
\end{equation}

\medskip

This formulation links national input–output systems into a single global structure where intermediate and final demands are endogenously allocated among competing suppliers. The resulting MRIO model can be expressed as:

\begin{equation}
x = (I - T   A)^{-1}(T   fd),
\label{eq:mrio_core}
\end{equation}
where $A$ is the block-diagonal matrix of total technical coefficients, $fd$ the vector of final demand, and $T$ the bilateral distribution of trade flows.

\begin{equation}
A =
\begin{pmatrix}
A_1 & 0 & \cdots & 0\\
0 & A_2 & \cdots & 0\\
\vdots & \vdots & \ddots & \vdots\\
0 & 0 & \cdots & A_N
\end{pmatrix}.
\label{eq:A}
\end{equation}

Traditional MRIO simulations treat the Leontief inverse $(I - T A)^{-1}$ as fixed, implying constant technical coefficients and trade relationships. This rigidity limits analysis when shocks alter prices or competitiveness. To overcome this, our model endogenizes the bilateral trade coefficients in $T$, allowing exporters’ market shares to adjust dynamically with relative prices and competitiveness.

\medskip

Following \citet{eaton_technology_2002}, delivered costs of supplying goods from $o$ to $d$ are:

\begin{equation}
c_{d,o} = \frac{c_o}{z_o} d_{d,o},
\label{eq:EK_cost}
\end{equation}
where $c_o$ is the wage in country $o$, $z_o$ its productivity, and $d_{d,o}>1$ the bilateral trade cost.

\medskip

We simplify the trade structure by adopting an Armington specification, where goods from different origins are imperfect substitutes. Let $X_{d,o}$ denote the expenditure by country $d$ on goods from $o$, $X_d=\sum_o X_{d,o}$ total expenditure, and $c_{d,o}$ the delivered import price. Under Armington preferences, the bilateral trade share is:

\begin{equation}
\frac{X_{d,o}}{X_d} =
\frac{a_o^\sigma (c_{d,o} d_{d,o})^{1-\sigma}}
{\sum_{k=1}^{N} a_k^\sigma (c_{d,k} d_{d,k})^{1-\sigma}}.
\label{eq:armington}
\end{equation}

where $a_k$ represents the preference parameter and $\sigma$ represents the elasticity of substitution between different sources.
\medskip

Combining all components yields a global MRIO model linking production, demand, and prices across economies. The system determines output, income, domestic and import prices, and bilateral flows through interdependent equations:

\begin{equation}
(I - T A)x = T fd,
\label{eq:production}
\end{equation}

This equation defines the level of production ($x$) in each country and sector as a function of final demand ($fd$), intermediate inputs and the trade shares ($T$).

\begin{equation}
fd = \psi(p),
\label{eq:demand}
\end{equation}

The final demand is expressed as a function of domestic prices ($p$). This makes the components of aggregate demand (consumption, investment and public expenditure) endogenous.

\begin{equation}
\Delta p = (I - T A')^{-1} \Delta p_m,
\label{eq:prices}
\end{equation}

Changes in domestic prices are determined by two factors: the cost matrix $(I - T A')^{-1}$ and variations in import prices.

\begin{equation}
T = \gamma(p),
\label{eq:trade_shares}
\end{equation}

The trade share matrix ($T$) is determined endogenously, depending on the relative prices.

\begin{equation}
m = [u' T_2 (A \hat{x} + fd)],
\label{eq:imports}
\end{equation}

Aggregating across all sectors and regions yields the total import vector $m$, which represents the demand side of the external balance. $T_2$ is the off-diagonal matrix of $T$ and $u$ is a vector of ones.

\begin{equation}
e = [T_2 (A \hat{x} + fd) u],
\label{eq:exports}
\end{equation}

Exports ($e$) are calculated as the product of the import-coefficient matrix ($T_2$) and total demand $(A \hat{x} + fd)$. This ensures bilateral consistency between global imports and exports in all economies.

\subsection{Data}
\label{subsec:data}

This section describes the data used to assess the employment impact of the ongoing trade conflict. The International Labour Organization (ILO) compiles household survey microdata collected by national statistical offices from 161 different countries. The raw datasets are harmonised according to international standards to produce comparable indicators, providing a comprehensive and internationally consistent set of labour statistics.

\medskip

All employment-related parameters in this study are obtained from a satellite account derived from ILOSTAT employment data. This classification includes the working-age population actively employed, defined as individuals engaged in any activity for at least one hour to produce goods or services for pay or profit, even if temporarily absent from work \citep{ILO2013ICLS19}.

\medskip

For the calculation of the model, we use the OECD’s Inter-Country Input–Output (ICIO) matrix with base year 2019. The ICIO database synthesises information on 45 economic activities for 76 individual countries and one additional region representing the rest of the world (RoW). The input–output table represents the monetary transactions between productive and institutional sectors, capturing the sectoral composition of GDP components, such as household consumption, government expenditure, investment, exports, and imports, as well as the breakdown of taxes and value added (wages and profits).

\medskip

In the MRIO version, the intermediate transactions matrix $Z$ contains the purchases and sales of inputs between sectors within the same country and across external destinations. Final demand, by contrast, is a matrix whose rows contain the final purchases made by countries at the sectoral level, including imports from other origins. Its columns detail the components of domestic demand: private consumption, public consumption, and investment. Finally, because a country's imports correspond to the exports of other countries, external sales at the sectoral level are represented in the final demand matrix in the off-diagonal columns.

\clearpage
\section{Policy Scenarios}
\label{sec:scenarios}

The next stage of the analysis involves generating counterfactual scenarios encompassing policy changes to be evaluated. A policy change is modelled as a shock to an exogenous variable in the system. Once the shock is applied, the model is solved for a new equilibrium. If a satisfactory solution is achieved, the results are compared with those of the baseline scenario to assess the implications of each policy configuration.

\medskip

Alternative scenarios may reflect fiscal incentives, structural shifts in demand, or policies promoting technology transfer. In this paper, policy shocks are derived from the recent trade conflict between the United States and its major partners. The objective is to estimate long-term effects and assess the risks of protectionist measures on trade, employment, economic integration, and globalisation.

\medskip

As shown in Table~\ref{tab:scenarios}, three scenarios are developed to assess the impact of tariff escalation.

\begin{table}[htbp]
\centering
\caption{Scenarios to consider for measuring the impact of trade tensions}
\label{tab:scenarios}
\begin{threeparttable}
\small
\setlength{\tabcolsep}{4.5pt}
\renewcommand{\arraystretch}{1.1}
\begin{tabular}{p{3cm} p{3.8cm} p{3.8cm} p{3.8cm}}
\toprule
\textbf{Actor} & \textbf{Scenario 1} & \textbf{Scenario 2} & \textbf{Scenario 3} \\
\midrule
\textbf{United States} &
Imposes an import tariff on all partners and additional increases by sector (e.g., steel, cars)---see Table~\ref{tab:chronology}. &
Imposes an import tariff on all partners and additional sectoral increases (e.g., steel, cars) including updates and agreements in force until August. &
Reduces the general increase in import tariffs to 10\%; maintains a 30\% tariff on imports from China. \\[4pt]

\textbf{Trading partners} &
No actions. &
Retaliation by Brazil, Canada, and China. &
Developing (non–high-income) countries grant a 50\% reduction in import duties on goods from the United States; China imposes a 10\% tariff on imported products from the United States \\

\bottomrule
\end{tabular}
\begin{tablenotes}
\footnotesize
\item \textit{Source:} Authors’ elaboration.
\end{tablenotes}
\end{threeparttable}
\end{table}

\paragraph{Scenario 1: Generalised tariff escalation.}
The United States imposes a uniform import tariff on all trading partners, in line with the measures announced during the April “Liberation Day” declaration. This represents a hypothetical generalized protectionist escalation, affecting all imports. Sector-specific tariff increases are also included, such as a 50 per cent  rise on steel and aluminium imports for all partners—except the United Kingdom, which faces a 25 per cent  increase following negotiations. This setting captures the initial phase of global trade tensions.

\medskip

\paragraph{Scenario 2: Updated tariffs with bilateral adjustments.}
This scenario updates the tariffs in force until August, accounting for subsequent bilateral negotiations and retaliatory actions. In several sectors, tariff increases exceed those imposed unilaterally. However, the United Kingdom, Japan, the European Union and Vietnam secure partial exclusions or reductions. For example, Vietnam’s tariff rate falls from 46 per cent  to 20 per cent, while the U.K. retains a 25 per cent  rate on steel and aluminium. China, Brasil and Canada retaliate imposing tariffs on United States imports. 

\medskip

\paragraph{Scenario 3: Partial de-escalation.}
This scenario assumes a partial rollback of the tariff escalation. The United States limits its general import tariff to 10 per cent  but maintains a 30 per cent  rate on imports from China. In retaliation, developing (non–high-income) countries reduce import duties on United States goods by 50 per cent, seeking to mitigate the trade war’s impact. China, however, imposes a 10 per cent  tariff on imports from the United States. This configuration reflects a mixed outcome—partial easing combined with persistent bilateral tensions between the two largest economies.

\clearpage
\section{Results}
\label{sec:results}

In consideration of the scenarios presented and the methodology developed in the preceding section, this section presents the results of the simulation, with a primary focus on employment and export dynamics for the countries and productive sectors involved in the analysis.

\medskip

As expected, the overall impact on employment is negative, resulting from the increase in tariffs and the associated changes in relative prices. In Table~\ref{tab:employment}, the data illustrate changes in both percentage terms and absolute values of employment, with the analysis conducted by income level and region of the countries within the model. Following the World Bank's classification\footnote{\url{https://blogs.worldbank.org/en/opendata/world-bank-country-classifications-by-income-level-for-2024-2025}}, a distinction is made between Asian countries and other upper middle-income countries, and between the European Union and other high-income countries.

\medskip

In Scenario~1, the most significant percentage changes in employment occur in high-income rest (HIC-Rest), upper-middle-income Asia, and low-income Asia, with --1.4 per cent, --1.31 per cent, and --1.26 per cent, respectively. For China and the United States, the percentage change is negative in both cases, with the impact being greater for China (--1.23 per cent) than for the United States (--0.15 per cent). In absolute terms, China experiences the largest decline, with more than 6.3~million jobs lost. The low-income Asia region loses about 6.1~million jobs, followed by the upper-middle-rest income group with 2.9~million jobs lost.

\medskip

In Scenario~2, which incorporates bilateral agreements between the United States and other partners, percentage changes tend to decrease slightly but with greater absolute job losses. The United States faces a 0.9 per cent  decline, while the HIC-Rest and upper-middle-income Asia groups record the largest employment declines at 1.4 per cent  and 1.3 per cent, respectively. Low-income Asia (6.6~million jobs), China (6.1~million jobs), and upper-middle-rest (3.4~million jobs) continue to experience the most severe losses.

\medskip

In Scenario~3, where the United States raises tariffs by only 10 per cent  (except China), the differential impact across regions increases. China suffers the largest job loss, exceeding 6.9~million, followed by low-income Asia (2.3~million) and upper-middle-rest (1.9~million). In general, the results for all three scenarios reveal that escalating trade tensions cause widespread employment losses, with the most severe effects concentrated in developing Asian economies and China. Even under partial de-escalation, the negative global employment effects remain substantial.

\medskip

\begin{table}[htbp]
\centering
\caption{Impact on employment in Scenarios~1, 2, and~3 (thousand jobs and percentage change)}
\label{tab:employment}
\begin{threeparttable}
\small
\setlength{\tabcolsep}{5pt}
\renewcommand{\arraystretch}{1.1}
\begin{tabular}{lrrrrrr}
\toprule
\textbf{Income group} & \multicolumn{2}{c}{\textbf{Scenario~1}} & \multicolumn{2}{c}{\textbf{Scenario~2}} & \multicolumn{2}{c}{\textbf{Scenario~3}} \\
\cmidrule(lr){2-3} \cmidrule(lr){4-5} \cmidrule(lr){6-7}
 & \textbf{Thousands} & \textbf{Change} & \textbf{Thousands} & \textbf{Change} & \textbf{Thousands} & \textbf{Change} \\
\midrule
China & -6{,}322 & -1.23\% & -6{,}183 & -1.22\% & -6{,}972 & -1.29\% \\
HIC-Asia & -860 & -0.85\% & -798 & -0.77\% & -549 & -0.62\% \\
HIC-EU27 & -798 & -0.60\% & -788 & -0.58\% & -697 & -0.57\% \\
HIC-Rest & -944 & -1.40\% & -1{,}012 & -1.39\% & -596 & -0.89\% \\
Low income & -195 & -0.41\% & -185 & -0.38\% & -134 & -0.27\% \\
Low income-Asia & -6{,}186 & -1.26\% & -6{,}675 & -1.09\% & -2{,}341 & -0.47\% \\
USA & -846 & -0.15\% & -1{,}330 & -0.91\% & -541 & -0.12\% \\
Upper middle-Asia & -1{,}612 & -1.31\% & -1{,}601 & -1.28\% & -594 & -0.60\% \\
Upper middle-Rest & -2{,}903 & -0.93\% & -3{,}405 & -1.07\% & -1{,}888 & -0.69\% \\
Other regions & -1{,}037 & -0.28\% & -1{,}033 & -0.27\% & -1{,}026 & -0.25\% \\
\midrule
\textbf{Total} & -21{,}702 & & -23{,}010 & & -15{,}336 & \\
\bottomrule
\end{tabular}
\begin{tablenotes}
\footnotesize
\item \textit{Source:} Authors’ estimation.
\end{tablenotes}
\end{threeparttable}
\end{table}

\medskip

Table~\ref{tab:exports} summarises the impact of the trade conflict on exports. The United States exhibits the largest decline in external sales in Scenario~1, falling by 11 per cent, followed by China and the HIC-Rest group with drops of 7 per cent  and 5.6 per cent, respectively. In absolute terms, China suffers the greatest decline (USD~141{,}774~million), followed by the HIC-EU27 (USD~109{,}882~million) and the United States (USD~108{,}616~million).

\medskip

The pattern persists in Scenario~2, where the United States becomes the most affected economy, facing an export contraction of 17.4 per cent. China and low-income Asia follow with a 7 per cent  and 5.9 per cent  fall, respectively. In Scenario~3, export losses moderate, yet remain significant: the United States (–9.7 per cent), China (–5.9 per cent), and HIC-Rest (–3.6 per cent) lead the declines. Aggregate global exports fall by USD~596{,}463~million, representing a 32.4 per cent  reduction relative to Scenario~2.

\medskip

\begin{table}[htbp]
\centering
\caption{Impact on exports in Scenarios~1, 2, and~3 (USD~million and percentage change)}
\label{tab:exports}
\begin{threeparttable}
\small
\setlength{\tabcolsep}{5pt}
\renewcommand{\arraystretch}{1.1}
\begin{tabular}{lrrrrrr}
\toprule
\textbf{Income group} & \multicolumn{2}{c}{\textbf{Scenario~1}} & \multicolumn{2}{c}{\textbf{Scenario~2}} & \multicolumn{2}{c}{\textbf{Scenario~3}} \\
\cmidrule(lr){2-3} \cmidrule(lr){4-5} \cmidrule(lr){6-7}
 & \textbf{Value} & \textbf{Change} & \textbf{Value} & \textbf{Change} & \textbf{Value} & \textbf{Change} \\
\midrule
China & -141{,}774 & -7.0\% & -143{,}233 & -7.0\% & -120{,}809 & -5.9\% \\
HIC-Asia & -106{,}634 & -4.5\% & -97{,}952 & -4.0\% & -67{,}200 & -3.0\% \\
HIC-EU27 & -109{,}882 & -1.6\% & -107{,}888 & -1.5\% & -101{,}701 & -1.6\% \\
HIC-Rest & -124{,}104 & -5.6\% & -132{,}300 & -5.8\% & -78{,}473 & -3.6\% \\
Low income & -2{,}895 & -2.1\% & -2{,}813 & -2.0\% & -2{,}244 & -1.6\% \\
Low income-Asia & -61{,}833 & -6.1\% & -62{,}231 & -5.9\% & -29{,}640 & -2.5\% \\
USA & -108{,}616 & -11.0\% & -177{,}199 & -17.4\% & -95{,}625 & -9.7\% \\
Upper middle-Asia & -26{,}185 & -5.0\% & -25{,}997 & -4.9\% & -12{,}079 & -2.1\% \\
Upper middle-Rest & -108{,}572 & -3.8\% & -118{,}349 & -4.8\% & -74{,}424 & -3.1\% \\
Other regions & -16{,}653 & -1.7\% & -15{,}417 & -1.7\% & -14{,}268 & -1.8\% \\
\midrule
\textbf{Total} & -807{,}148 & & -883{,}380 & & -596{,}463 & \\
\bottomrule
\end{tabular}
\begin{tablenotes}
\footnotesize
\item \textit{Source:} Authors’ estimation.
\end{tablenotes}
\end{threeparttable}
\end{table}

\medskip

Table~\ref{tab:topcountries} identifies the 15 countries experiencing the most significant employment losses in all scenarios. In Scenario~1, China, India and Mexico record the largest declines—6.3, 2.9, and 2.2~million jobs, respectively. Percentage-wise, Canada (--6.8 per cent), Mexico (--6.0 per cent), and Viet~Nam (--3.6 per cent) show the greatest contractions. Other Asian economies, including India, Indonesia, Bangladesh, Thailand, and South Korea, also show highly negative results.

\medskip

In Scenario~2, China and India remain the most affected countries in absolute terms, while Brazil (–2.1 per cent) suffers additional losses due to retaliatory measures. Canada continues to exhibit the steepest proportional decline (–7.2 per cent). The United States records over one million job losses (–0.9 per cent), higher than in Scenario~1.

\medskip

In Scenario~3, China again registers the most substantial job loss (6.9~million), followed by India and Mexico. Canada (--3.2 per cent) and Mexico (--3.6 per cent) remain the most negatively affected in percentage terms, while the United States shows a limited decline (–0.1 per cent) due to reduced retaliation and partial tariff easing.

\medskip

\begin{table}[htbp]
\centering
\caption{Top 15 countries most affected in employment (thousand jobs and percentage change)}
\label{tab:topcountries}
\begin{threeparttable}
\scriptsize
\renewcommand{\arraystretch}{1.3}
\setlength{\tabcolsep}{4pt}

\begin{tabular}{lrr lrr lrr}
\toprule
\multicolumn{3}{c}{\textbf{Scenario 1}} & \multicolumn{3}{c}{\textbf{Scenario 2}} & \multicolumn{3}{c}{\textbf{Scenario 3}} \\
\cmidrule(lr){1-3} \cmidrule(lr){4-6} \cmidrule(lr){7-9}
\textbf{Country} & \textbf{Thousands} & \textbf{Change} &
\textbf{Country} & \textbf{Thousands} & \textbf{Change} &
\textbf{Country} & \textbf{Thousands} & \textbf{Change} \\
\midrule
China        & -6{,}322 & -1.23\% & China        & -6{,}183 & -1.22\% & China        & -6{,}972 & -1.29\% \\
India        & -2{,}943 & -1.42\% & India        & -4{,}479 & -1.94\% & India        & -1{,}521 & -0.88\% \\
Mexico       & -2{,}279 & -6.01\% & Mexico       & -2{,}330 & -6.12\% & Mexico       & -1{,}329 & -3.66\% \\
Viet Nam     & -1{,}984 & -3.59\% & United States& -1{,}330 & -0.91\% & RoW          & -1{,}026 & -0.25\% \\
RoW          & -1{,}037 & -0.28\% & RoW          & -1{,}033 & -0.27\% & United States&   -541   & -0.12\% \\
Indonesia    &   -975   & -0.80\% & Indonesia    &   -969   & -0.78\% & Viet Nam     &   -459   & -1.05\% \\
United States&   -846   & -0.15\% & Viet Nam     &   -938   & -1.94\% & Indonesia    &   -313   & -0.30\% \\
Canada       &   -559   & -6.77\% & Brazil       &   -671   & -2.13\% & Japan        &   -279   & -0.71\% \\
Bangladesh   &   -524   & -0.45\% & Canada       &   -649   & -7.18\% & Canada       &   -271   & -3.25\% \\
Thailand     &   -479   & -1.67\% & Bangladesh   &   -535   & -0.45\% & Korea        &   -204   & -1.06\% \\
Japan        &   -453   & -1.19\% & Thailand     &   -471   & -1.63\% & Brazil       &   -197   & -0.55\% \\
Korea        &   -325   & -1.72\% & Japan        &   -393   & -0.87\% & Germany      &   -195   & -0.78\% \\
Pakistan     &   -242   & -0.26\% & Korea        &   -324   & -1.69\% & Thailand     &   -187   & -0.66\% \\
Germany      &   -236   & -0.87\% & Pakistan     &   -247   & -0.25\% & Philippines  &   -117   & -0.59\% \\
Brazil       &   -216   & -0.55\% & Germany      &   -235   & -0.84\% & Russia       &   -117   & -0.23\% \\
\midrule
\textbf{Sub-total} & -19{,}420 &  & \textbf{Sub-total} & -20{,}787 &  & \textbf{Sub-total} & -13{,}729 &  \\
\textbf{Total}     & -21{,}702 &  & \textbf{Total}     & -23{,}010 &  & \textbf{Total}     & -15{,}610 &  \\
\bottomrule
\end{tabular}

\begin{tablenotes}
\footnotesize
\item \textit{Source:} Authors’ estimation.
\end{tablenotes}
\end{threeparttable}
\end{table}

\medskip
The analysis of the sectors affected, according to the ICIO classification, is presented in Table~\ref{tab:sectors}, which shows the top 15 sectors in terms of job losses. In the initial scenario, the {Agriculture} (5.22~million jobs), {Retail trade} (3.43~million), {Textiles} (2.46~million) and {Manufacturing nec.} sectors (1.26~million) show the most significant loss in absolute terms. However, the largest percentage changes are observed in {Motor vehicles} (–2.0 per cent), {Plastics products} (–1.6 per cent) and {Computer and electronic products} (–1.4 per cent).

\medskip

In Scenario~2, the sectors most affected in terms of job losses remain largely the same as in Scenario~1, with only slight differences in percentage changes. The sectors showing the most substantial relative declines are again {Motor vehicles} (–2.1 per cent), {Plastic products} (–1.5 per cent), and {Computer and electronic products} (–1.5 per cent).

\medskip

In Scenario~3, the {Agricultural} sector exhibits a decrease of 3.3~million jobs—almost 1.7~million fewer than in the previous scenarios—due to the easing of trade tensions. However, the {Trade} sector continues to experience job losses exceeding 3.1~million. The {Textiles} sector also remains among the most affected, with over 1.15~million jobs lost. In relative terms, the {Basic metals} and {Motor vehicles} sectors show notable contractions of –1.8 per cent  and –1.35 per cent, respectively.

\clearpage
% -----------------------------------------------
\begin{sidewaystable}[p]
\centering
\caption{Top 15 sectors most affected in employment (thousand jobs and percentage change)}
\label{tab:sectors}
\begin{threeparttable}
\scriptsize
\renewcommand{\arraystretch}{1.1}
\setlength{\tabcolsep}{4pt}

\begin{tabular}{lrr lrr lrr}
\toprule
\multicolumn{3}{c}{\textbf{Scenario 1}} & \multicolumn{3}{c}{\textbf{Scenario 2}} & \multicolumn{3}{c}{\textbf{Scenario 3}} \\
\cmidrule(lr){1-3} \cmidrule(lr){4-6} \cmidrule(lr){7-9}
\textbf{Sector (ICIO)} & \textbf{Thousands} & \textbf{Change} &
\textbf{Sector (ICIO)} & \textbf{Thousands} & \textbf{Change} &
\textbf{Sector (ICIO)} & \textbf{Thousands} & \textbf{Change} \\
\midrule
Agriculture                                & -5{,}225 & -0.67\% & Agriculture                                & -5{,}149 & -0.65\% & Agriculture                                & -3{,}319 & -0.46\% \\
Retail trade                               & -3{,}436 & -0.66\% & Retail trade                               & -3{,}641 & -0.66\% & Retail trade                               & -3{,}092 & -0.42\% \\
Textiles                                   & -2{,}468 & -1.30\% & Textiles                                   & -2{,}609 & -1.29\% & Textiles                                   & -1{,}151 & -0.55\% \\
Manufacturing n.e.c.                       & -1{,}267 & -1.38\% & Manufacturing n.e.c.                       & -1{,}495 & -1.33\% & Manufacturing n.e.c.                       &   -725   & -0.75\% \\
Motor vehicles                             &   -724   & -1.99\% & Motor vehicles                             &   -776   & -2.11\% & Land transport  &   -554   & -0.38\% \\
Land transport &   -678   & -0.60\% & Land transport &   -730   & -0.60\% & Motor vehicles                             &   -500   & -1.35\% \\
Computer and electronic           &   -615   & -1.43\% & Computer and electronic           &   -652   & -1.50\% & Computer and electronic                   &   -442   & -1.00\% \\
Metal products                             &   -524   & -1.28\% & Construction                               &   -628   & -0.22\% & Construction                               &   -404   & -0.15\% \\
Wood                                       &   -524   & -1.34\% & Wood                                       &   -588   & -1.27\% & Fabricated metal                  &   -359   & -0.80\% \\
Construction                               &   -498   & -0.19\% & Fabricated metal                  &   -540   & -1.23\% & Professional services     &   -322   & -0.39\% \\
Professional services     &   -408   & -0.54\% & Professional services     &   -458   & -0.55\% & Accommodation and food  &   -303   & -0.13\% \\
Administrative services        &   -387   & -0.73\% & Administrative services        &   -427   & -0.74\% & Wood                                       &   -299   & -0.80\% \\
Rubber and plastics                &   -353   & -1.59\% & Food and beverages                &   -379   & -0.52\% & Administrative services        &   -293   & -0.58\% \\
Food  and beverages                &   -351   & -0.52\% & Rubber and plastics               &   -351   & -1.53\% & Electrical equipment                       &   -253   & -0.58\% \\
Electrical equipment                       &   -327   & -0.99\% & Electrical equipment                       &   -349   & -1.02\% & Basic metals                               &   -251   & -1.87\% \\
\midrule
\textbf{Sub-total} & -17{,}784 &  & \textbf{Sub-total} & -18{,}772 &  & \textbf{Sub-total} & -12{,}267 &  \\
\textbf{Total}     & -21{,}702 &  & \textbf{Total}     & -23{,}010 &  & \textbf{Total}     & -15{,}193 &  \\
\bottomrule
\end{tabular}

\begin{tablenotes}
\footnotesize
\item \textit{Source:} Authors’ estimation.
\end{tablenotes}
\end{threeparttable}
\end{sidewaystable}

\clearpage

\medskip
Table~\ref{tab:labour_groups} shows the observed composition of job losses across different groups of employment: gender, age, skill, and formality status. The results reveal that the composition of job losses remains remarkably stable across the three scenarios, suggesting that the structure of vulnerability in global labour markets is persistent and largely determined by sectoral characteristics rather than the specific configuration of tariff retaliations.

\medskip
Across all scenarios, informal and unskilled workers are the  most adversely affected group by employment losses. Informal employment accounts for between 53 per cent  and 57 per cent  of total job losses, highlighting the disproportionate exposure of workers lacking formal protections and access to social safety nets. In addition, unskilled workers represent more than 80 per cent  of the total employment decline.

\medskip
From a demographic perspective, adult workers experience the greatest job losses, ranging from 85 per cent  to 87 per cent  of the total, consistent with their higher participation in sectors directly affected by trade disruptions. The gender distribution shows a modest yet persistent imbalance, with male workers accounting for approximately 65 per cent  of total job losses—aligned with their predominance in tradable, export-oriented sectors such as agriculture, metals, machinery, and transport equipment.

\medskip
\begin{table}[htbp]
\centering
\caption{Job losses distribution by labour group (\%)}
\label{tab:labour_groups}
\begin{threeparttable}
\small
\renewcommand{\arraystretch}{1.1}
\setlength{\tabcolsep}{6pt}

\begin{tabular}{lccc}
\toprule
\textbf{Group} & \textbf{Scenario 1} & \textbf{Scenario 2} & \textbf{Scenario 3} \\
\midrule
Formal      & 42.55 & 42.29 & 46.95 \\
Informal    & 57.45 & 57.71 & 53.05 \\
Skilled     & 17.50 & 18.75 & 19.79 \\
Unskilled   & 82.50 & 81.25 & 80.21 \\
Adult       & 85.91 & 85.92 & 87.14 \\
Youth       & 14.09 & 14.08 & 12.86 \\
Male        & 64.29 & 66.66 & 66.51 \\
Female      & 35.71 & 33.34 & 33.49 \\
\bottomrule
\end{tabular}

\begin{tablenotes}
\footnotesize
\item \textit{Source:} Authors’ estimation.
\end{tablenotes}
\end{threeparttable}
\end{table}

\clearpage
\section{Concluding remarks}
\label{sec:conclusion}

The simulations conducted in this paper confirm that the renewed escalation of trade tensions has a globally contractionary and regressive impact on employment, with job losses spreading across income groups, sectors, and worker types. The results show that no major economy is immune from these effects: even high-income countries, which often rely on diversified export portfolios, experience non-negligible declines in employment, while lower-income economies and upper middle-income exporters, such as China and India, bear the largest absolute losses. The findings emphasise that trade protectionism, whether or not amplified by retaliation, produces widespread welfare losses through multiple channels---reduced export competitiveness, higher import prices, and shrinking final demand---resulting in cumulative declines in both output and employment across the global production network.

\medskip
At the sectoral level, the results indicate that labour-intensive and trade-dependent industries such as agriculture, textiles, trade, motor vehicles, and transport are disproportionately affected. These sectors not only experience the largest absolute declines in employment but also exhibit the steepest percentage contractions, revealing the sensitivity of traditional industries to changes in global demand and intermediate input costs. Moreover, the persistence of results across scenarios, regardless of whether retaliation is introduced, highlights the structural nature of exposure embedded in global value chains. Tariff adjustments trigger cascading effects along supply value chains that cannot be fully offset by trade diversion, confirming that protectionist measures in one country impose systemic costs on others.

\medskip
A final consideration relates to foreign direct investment, which plays a critical role in the trade--employment nexus. The stated objective of the United States administration’s tariff policy is to attract investment into the United States, encouraging firms to relocate production domestically to avoid trade barriers and thereby create jobs. While this rationale aligns with the political narrative of restoring industrial employment, the results presented here suggest that the macroeconomic contraction and export losses triggered by protectionism may counteract these objectives. In practice, a global downturn in demand and rising production costs could deter new investment and slow capital reallocation. Future research should therefore investigate the interaction between tariffs, FDI flows, and employment dynamics to assess whether protectionist policies truly induce productive investment or simply redistribute global losses across sectors and regions.

\medskip
From a labour-market perspective, the analysis reveals a clear inequality dimension of trade shocks. The distribution of job losses across groups demonstrates that informal, unskilled, and male workers in tradable sectors are the most affected. This pattern points to the need for labour and social policies that address the asymmetric impact of global trade disruptions, particularly those targeting vulnerable groups lacking protection mechanisms. Future research should expand this framework to incorporate wage and income effects, occupational reallocation, and transition dynamics between sectors, as well as the role of labour policies, retraining programmes, and industrial strategies in mitigating employment risks.

\clearpage

\bibliographystyle{apalike} 
\bibliography{Trade_war.bib}

\clearpage

\appendix
\section{Simulated import tariffs} 

\begin{table}[htbp]
\centering
\resizebox{\textwidth}{!}{%
\begin{threeparttable}
\caption{Tariff Shocks by Country and Scenario}
\begin{tabular}{llcccccc}
\toprule
\multirow{2}{*}{Country} &
\multirow{2}{*}{Income--Region} &
\multicolumn{2}{c}{Scenario 1} &
\multicolumn{2}{c}{Scenario 2} &
\multicolumn{2}{c}{Scenario 3} \\
\cmidrule(lr){3-4} \cmidrule(lr){5-6} \cmidrule(lr){7-8}
& & USA $\uparrow$ Tariff & Retaliation 
& USA $\uparrow$ Tariff & Retaliation 
& USA $\uparrow$ Tariff & Retaliation \\
\midrule
Bangladesh         & Low income--Asia        & 37\% &     & 37\% &     & 10\% &     \\
Brazil             & Upper middle--Rest      & 50\% &     & 50\% & 50\% & 10\% &     \\
Brunei Darussalam  & HIC--Asia               & 24\% &     & 24\% &     & 10\% &     \\
Canada             & HIC--Rest               & 35\% &     & 35\% & 25\% & 10\% &     \\
Switzerland        & HIC--Rest               & 31\% &     & 31\% &     & 10\% &     \\
Chile              & HIC--Rest               & 10\% &     & 10\% &     & 10\% &     \\
China              & China                   & 35\% &     & 30\% & 10\% & 30\% & 10\% \\
Côte d'Ivoire      & Low income              & 21\% &     & 21\% &     & 10\% &     \\
Cameroon           & Low income              & 11\% &     & 11\% &     & 10\% &     \\
Colombia           & Upper middle--Rest      & 10\% &     & 10\% &     & 10\% &     \\
Costa Rica         & Upper middle--Rest      & 10\% &     & 10\% &     & 10\% &     \\
Hong Kong, China   & HIC--Asia               & 34\% &     & 34\% &     & 10\% &     \\
Indonesia          & Upper middle--Asia      & 32\% &     & 32\% &     & 10\% &     \\
India              & Low income--Asia        & 51\% &     & 51\% &     & 10\% &     \\
Iceland            & HIC--Rest               & 10\% &     & 10\% &     & 10\% &     \\
Israel             & HIC--Rest               & 17\% &     & 17\% &     & 10\% &     \\
Jordan             & Low income              & 20\% &     & 20\% &     & 10\% &     \\
Japan              & HIC--Asia               & 24\% &     & 24\% &     & 10\% &     \\
Kazakhstan         & Upper middle--Rest      & 27\% &     & 27\% &     & 10\% &     \\
Cambodia           & Low income--Asia        & 49\% &     & 49\% &     & 10\% &     \\
Korea              & HIC--Asia               & 25\% &     & 25\% &     & 10\% &     \\
Lao PDR            & Low income--Asia        & 48\% &     & 48\% &     & 10\% &     \\
Mexico             & Upper middle--Rest      & 25\% &     & 25\% &     & 10\% &     \\
Myanmar            & Low income--Asia        & 44\% &     & 44\% &     & 10\% &     \\
Malaysia           & Upper middle--Asia      & 24\% &     & 24\% &     & 10\% &     \\
Nigeria            & Low income              & 14\% &     & 14\% &     & 10\% &     \\
Norway             & HIC--Rest               & 15\% &     & 15\% &     & 10\% &     \\
New Zealand        & HIC--Asia               & 10\% &     & 10\% &     & 10\% &     \\
Pakistan           & Low income--Asia        & 29\% &     & 29\% &     & 10\% &     \\
Philippines        & Low income--Asia        & 17\% &     & 17\% &     & 10\% &     \\
Thailand           & Upper middle--Asia      & 36\% &     & 36\% &     & 10\% &     \\
Tunisia            & Low income              & 28\% &     & 28\% &     & 10\% &     \\
Türkiye            & Upper middle--Rest      & 10\% &     & 10\% &     & 10\% &     \\
Chinese Taipei     & HIC--Asia               & 32\% &     & 32\% &     & 10\% &     \\
Viet Nam           & Low income--Asia        & 46\% &     & 20\% &     & 10\% &     \\
South Africa       & Upper middle--Rest      & 30\% &     & 30\% &     & 10\% &     \\
Rest of countries  & ROC                     & 10\% & --  & 10\% & --  & 10\% & --  \\
\bottomrule
\end{tabular}
\end{threeparttable}
}
\end{table}

\end{document}